# Supraventricular Tachycardia Detection and Classification Model of ECG signal Using Machine Learning

[1]Pampa Howladar, [2]Manodipan Sahoo

[1]IIEST, Shibpur, India , [2] IIT Dhanbad, India

*Abstract*— **Investigation on the electrocardiogram (ECG) signals is an essential way to diagnose heart disease since the ECG process is noninvasive and easy to use. This work presents a supraventricular arrhythmia prediction model consisting of a few stages, including filtering of noise, a unique collection of ECG characteristics, and automated learning classifying model to classify distinct types, depending on their severity. We de-trend and de-noise a signal to reduce noise to better determine functionality before extractions are performed. After that, we present one R-peak detection method and Q-S detection method as a part of necessary feature extraction. Next parameters are computed that correspond to these features. Using these characteristics, we have developed a classification model based on machine learning that can successfully categorize different types of supraventricular tachycardia. Our findings suggest that decision-tree-based models are the most efficient machine learning models for supraventricular tachycardia arrhythmia. Among all the machine learning models, this model most efficiently lowers the crucial signal misclassification of supraventricular tachycardia. Experimental results indicate satisfactory improvements and demonstrate a superior efficiency of the proposed approach with 97% accuracy.**

I. INTRODUCTION

Cardiac arrhythmias are cardiac rhythm variations that interrupt the heart's regular coordinated contraction pattern and decrease its effectiveness. Arrhythmia is a category of disorders where the heart beats too rapidly, too slowly or irregularly. Arrhythmias typically decrease hemodynamic effectiveness leading to circumstances in which the natural pacemaker of the heart produces an irregular rate or rhythm or normal conducting pathways are disrupted and rhythm regulates another region of the heart.

Supraventricular tachycardia (SVT) is an abnormally fast heartbeat caused by aberrant electrical activity in the upper chambers of the heart. Speeded-up rhythms, whether chronic or continuous, can cause fear in the patient, resulting in serious disease. In the absence of abnormal conduction, the ECG shows narrow, complicated tachycardia in this arrhythmia (e.g. bundle of a branch block) [1-4].

Symptoms and indications might emerge unexpectedly and heal on their own. Exercise, stress, and emotion can all contribute to normal or physiological fluctuations in heart rate, although they are seldom responsible for SVT. Episodes can last anywhere from a few minutes to a few days, and they frequently last until they are treated. The symptoms are characteristic with 150-270 beats per minute or more.

Over the previous few decades, there has been an extraordinary rate of surgical, clinical, and technological advancement. Since then, significant efforts have been undertaken to capitalize on technological breakthroughs and computer applications in the medical profession. Because the electrocardiogram (ECG) is used to examine the most essential organ in the human body, cardiologists are particularly interested in the most precise ECG analysis [5]. Several researchers in the field of ECG analysis have been conducted in an attempt to identify the signal at near-perfect speed automatically. Attempts have been undertaken over the last 30 years to replicate the skills of cardiologists and computer professionals. Many researchers that created many ECG identification and QRS detection methods that are widely recognized in the literature have tackled this topic. [6-10]. Among all methodologies, machine learning has received special attention due to its unique qualities such as nonlinearity, learning ability, and a universal approach to solving difficult signals such as QRS recognition, SVT diagnostics, and so on. This paper covers our methods for

developing and deploying a superior machine learning-based strategy for SVT categorization

*A. Electrocardiogram Signals:*

Electrocardiogram (ECG) recording is now used in the routine diagnosis or tracking of the effects of the heart or operating opioid rhythms that are not frequently asymptomatic. Recently, online signal processing, data reduction, and arrhythmia detection have been developed with microprocessor-based event recorders [11-12]. Microprocessor computing power helps one to mount digital filters for noise cancellation and arrhythmia identification [13][17]. Since contraction it reflects the electrical activity within the heart, the time and shape of occurrence give plenty of information about the heart's condition. A schematic record of a normal heartbeat, where the points P, Q, R, S, and T can be seen, is shown in Fig. 1. The ECG signals are usually measured in a range of ±2 mV and 0.05-150 Hz bandwidth.

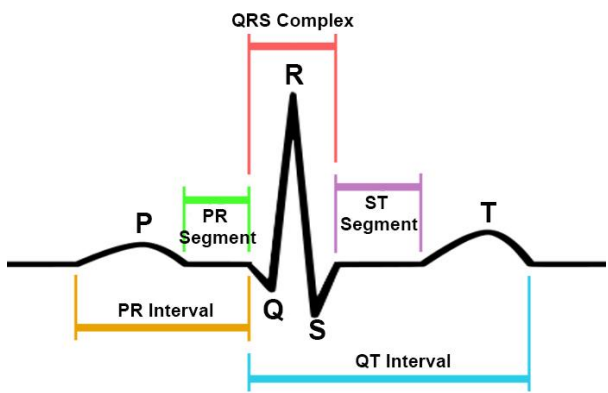

Fig 1: A diagram of a normal ECG signal

*B. Supraventricular Tachycardia Arrhythmias and its classification based on severity:*

Arrhythmias are heart abnormalities that produce irregular cardiac rhythms. Fig 2 depicts the categorization of cardiac arrhythmias. Rhythmic heart problems (heart arrhythmias) arise when electric pulses are not correctly coordinated, causing the heart to beat excessively quickly, slowly, or irregularly.

SVT is a rapid heart rhythm that comes from aberrant electricity in the upper region of the heart. "Supra" signifies the upper region of the heart while the below chambers of the heart mean "ventricular." In this type of arrhythmia, the ECG will display a narrow-complex tachycardia in the absence of aberrant conduction (e.g. bundle branch blocks). Most people with SVT enjoy a healthy life without limitations or therapy.

For others, changes in lifestyles, drugs, and cardiac procedures may be necessary to regulate or remove fast heartbeats and symptoms. There are four primary types of SVT: paroxysmal supraventricular tachycardia (PSVT), atrial fibrillation (AF), Wolff–Parkinson–White syndrome (WPW), and atrial flutter [14]. Of these four kinds, atrial and Wolff-Parkinson-White are dangerous and have a higher risk of heart arrest, heart failure, and stroke [15-16]. We define SVT as non-critical SVT, AF, and WPW according to severity.

*Symptoms:* Symptoms of Supraventricular Tachycardia include dyspnea, pressure or chest discomfort, light-headedness or dizziness, weariness, palpitations (including potential pulsations in the neck), chest pain (more severe than discomfort), and sudden death (may occur with Wolff-Parkinson-White syndrome).

*Treatment:* Medication, certain motions, catheter-based treatments (ablation), and an electrical shock to the heart (cardioversion) can all assist in slowing the heart.

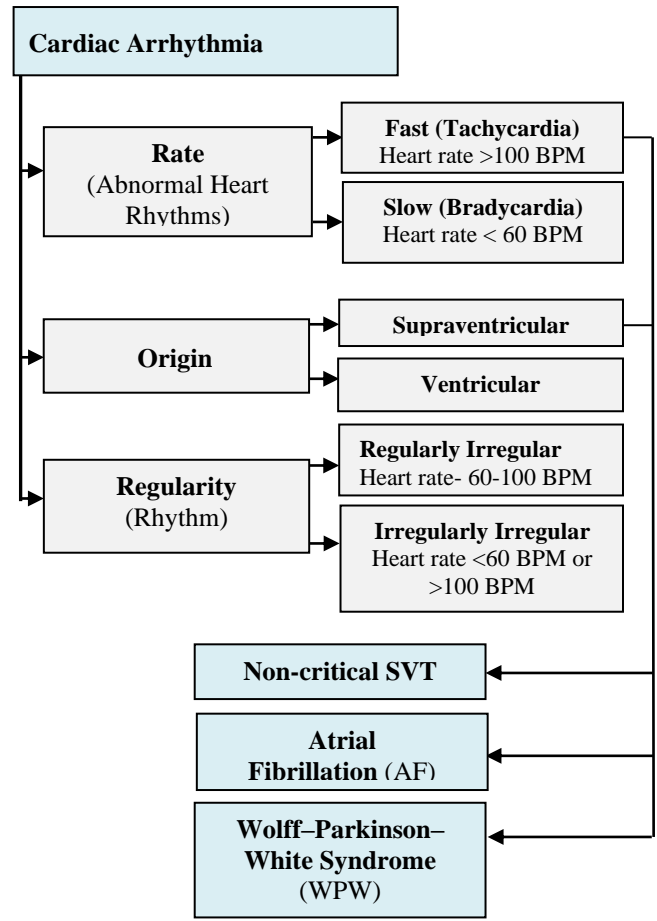

Fig 2: Cardiac arrhythmia classification

*C. Related Works:*

Numerous algorithms based on supraventricular tachycardia arrhythmia have been developed during these years. These include set of rules considered by optimal path forest [17], artificial neural networks [18-19], cardiologists [20-21], SVM [22-24], auto-regressive modeling [25], hidden Markov model (HMM) [26-28]. Although these methods have shown benefits in the diagnosis of supraventricular arrhythmia, they have some drawbacks. Some techniques are too difficult to enforce or compute, some do not differentiate between abnormal and normal situations which are usually not enough for action.

*D. Our Contributions:*

In this paper, we propose a Cardiac Arrhythmia classification model of supraventricular tachycardia. The long-term ECG monitoring is normal criterion for diagnosis of supraventricular and ventricular arrhythmia.
The paper has the following main contributions:
We de-noise and de-trend the signal to eliminate the noise during pre-processing, in order to better detect ECG features.
Thereafter necessary ECG features are extracted with help of our proposed methodology. After extracting features, necessary parameters related to supraventricular arrhythmia are calculated. These parameters are QRS duration, RR interval, PR interval, heartbeat rate (HBR), the standard deviation of the differences between successive RR intervals (SDSD) and Root mean square of the successive differences (RMSSD).
An effective machine learning-based classification model has been developed using these parameters in order to predict and correct the diagnosis of supraventricular arrhythmia detection. To the best of our understanding, for the first time various machine learning-based classification methods were assessed and a technique is chosen among these classification methods based on their high performance in order to diagnose the supraventricular and ventricular arrhythmias of ECG signal.
The remainder of this paper is also arranged accordingly. In Section II, our SVT detection method has been presented. The SVT classification method is explained in Section III. Results and discussion are presented in Section IV. Finally, in Section V, concluding remarks are discussed.

## II. SVT DETECTION METHOD

We have used one machine learning-based approach for SVT classification. The first approach is focused on heartbeat rate variation and morphological properties derived from the signals and uses machine learning algorithms based on those classification attributes. Fig 3 shows a flowchart with a brief overview of our proposed method. In this section, we will describe this flowchart in detail.

*A. Data Preprocessing:*

In order to eliminate noise artifacts, the ECG signal needs to be pre-processed. During pre-processing stage, A butterworth filter, cutoff frequencies at both 0.5 and 40 Hz, was added to raw ECG for removing noise at the signal pre-processing stage, whereby the baseline wandering was suppressed by a double median filters with orders of 0.2 and 0.6 times the sampling frequency.
Fig. 4 shows the ECG signal preprocessing stages before the signal is ready for feature extraction. The technique for removing the baseline signal drift [8-9] is called de-trending, and signal noise removal is known as a de-noising procedure. These two approaches fall under the field of ECG signal pre-processing. Fig. 5(a) shows the baseline drift of the record 102 signal, and the final signal is seen in Fig. 5 (b) after eliminating the baseline drift. Likewise Fig. 6(a) shows the noise presence of record 101 signal, and the final signal is seen in Fig. 6 (b) after eliminating drift and noise.

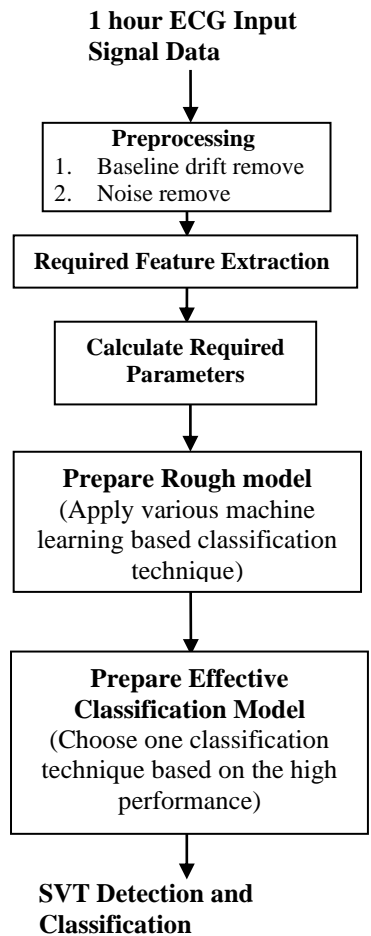

Fig 3: Flowchart of our proposed method

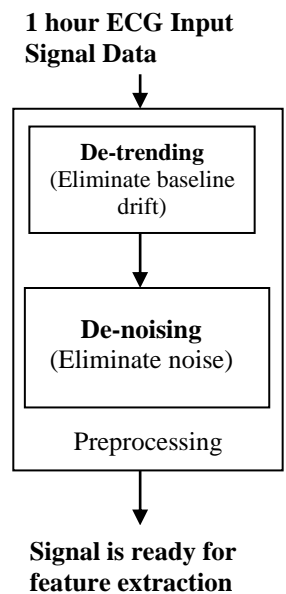

Fig 4: ECG Signal Preprocessing before feature extraction

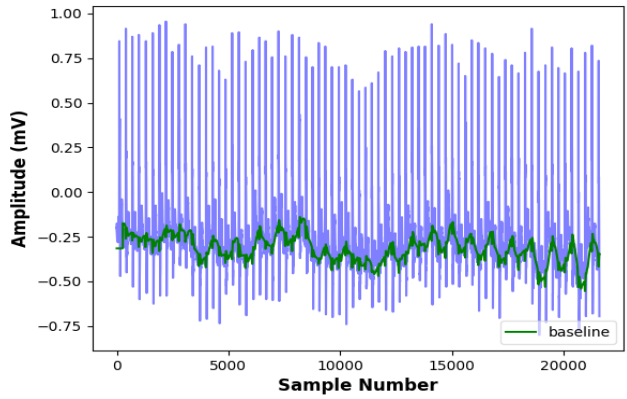

(a)

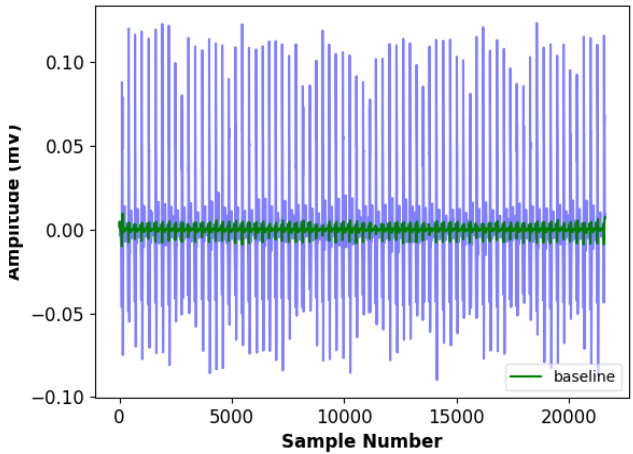

(b)

Fig 5: Signal baseline (a) with drift (b) after elimination of record 102

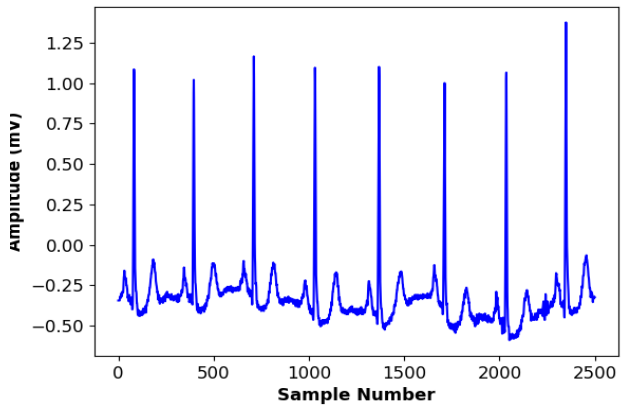

(a)

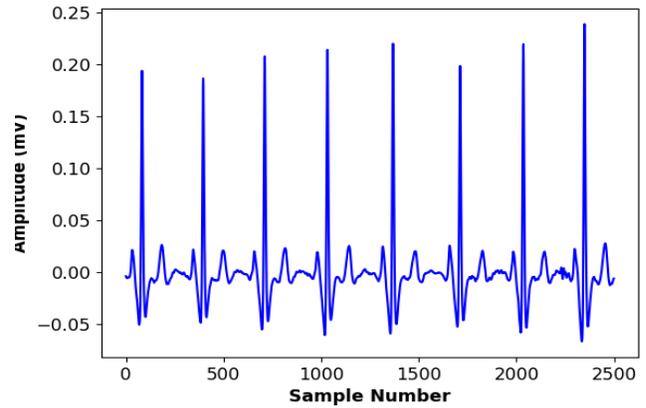

(b)

Fig 6: Signal (a) with drift and noise (b) after drift and noise elimination of record 101

*B. Feature extraction:*

The received signal is used to extract necessary characteristics (Q, R, and S). These characteristics are then utilized to calculate the required parameters. The criticality of the signal is determined by comparing these values to the ECG standard signal values. Table I shows required ECG features with acronyms.

Table I. Required ECG features with acronyms

| Abbreviations | Acronyms |
|---|---|
| **IBI** | inter-beat interval |
| **HBR** | Heart Beat Rate |
| **RMSSD** | Root mean square of the sucessive differences |
| **SDSD** | The standard deviation of the differences between successive RR intervals |

For supraventricular tachycardia detection using machine learning, the following parameters shown in Table 2 are considered here. Required ECG features with normal range are listed in Table II.

Table II. ECG features with normal range

| Features | Normal Range |
|---|---|
| RR interval | 0.6-1 s |
| RMSSD | 21-70 ms |
| SDSD | 141±39 ms |
| IBI | 600-900 ms |
| QRS duration | Upto 0.10s |
| HBR | 60-100 BPM |
| PR interval | 120-200 ms |

We proposed the R-peak detection algorithm in order to detect QRS complex as described in following steps of our proposed algorithm 1.
Algorithm 1 is briefed as follows:

**Step1:** Compute the total number of samples (S) of a given recorded signal.
**Step2:** Compute total time (t) with respect to the total number of samples.
**Step3:** Calculate time for each sample $t_S$ using (S/t).
**Step4:** Calculate the total number of samples which should be present between RR interval ($t_{RR}/t_S$) where $0.6 < t_{RR} < 1$.
**Step5:** Compute the total number of iterations (I) using (S/$S_{RR}$).
**Step6:** For each iterations store $S_{RR}$ number of data to Arr [$S_{RR}$] that sequentially collected from the total number of samples
**Step7:** For each sample point of Arr [$S_{RR}$] if the sample's new amplitude value is higher than the current amplitude value then it is considered as the R peak of the array. Repeat this step for every $S_{RR}$ number of samples.

**Algorithm 1**

**Input:** Recorded ECG Signal
**Assumption:** $0.6 < t_{RR} < 1$ (normal RR interval range)
**Output:** R-peak detection

**Procedure**
1. S ← Count total number of samples of recorded signal
2. t ← Calculate total time with respect to total number of samples
3. $t_S$ ← Calculate time for each sample (S/t)
4. $S_{RR}$ ← Calculate total number of samples which should be present between RR interval ($t_{RR}/t_S$)
5. I ← Calculate total number of iterations (S/$S_{RR}$)
6. **for** each iterations
7. **do** Arr [$S_{RR}$] ← Store $S_{RR}$ number of data sequentially collected from total number of samples
8.     for each sample point $O_s \in$ Arr [$S_{RR}$]
9.        Select and apply new operator
10.       Evaluate new state
11.       If sample's new amplitude value (mv) is higher than current amplitude value then it is considered as R peak of the array.
12. **end**

Once R-peak is detected we can easily calculate RR interval features.
Another feature RMSSD is the root mean square of the successive R-R interval difference. Equation 1 is the formula for calculating RMSSD where N is the total number of samples. The corresponding standard deviation of successive RR intervals (SDSD) shown in Equation 2 is a short-term variability only [29] and the IBI (interbeat interval) feature is assessed by the time interval between R-Waves.
Equation 1:

$$RMSSD = \sqrt{\frac{\sum_{i=1}^{N-1}(RR_i - RR_{i+1})^2}{N-1}}$$

Equation 2:
$RR_i - RR_{i+1} = D_i$
This variation is then used in order to find out the equation of SDSD.

$$SDSD = \sqrt{\frac{\sum_{i=1}^{i=n-1}(D_i - D_{mean})^2}{n-1}}$$

Where:
$i$ = interval index
$n$ = total number of intervals
$n-1$ = number of interval differences

$$D_{mean} = \frac{1}{n-1}\sum_{i=1}^{i=n-1} D_i$$

For QRS complex, an efficient technique for Q and S point detection is necessary. Algorithm 2 explains our technique elaborately.

**Algorithm2**

**Input:** R-peak value collected from algorithm 1 of a recorded signal
**Assumption:** S=total number of samples, $R_n$=Value [$Arr_{R\ peak}$ [n]]

**Output:** Q and S point detection

**Procedure**
1. Store R-peak value in an array $Arr_{R\ peak}$ [n]
2. C= Count array value
3. **while** $R_0 \in S$
4.    **for** 0 to $R_0$
5.      Evaluate initial state as current state
6.      Select and apply new operator
7.      If new sample's amplitude is less than current sample then it is marked as current state.
8.      The latest amplitude value of current state is marked as $Q_0$ point of QRS.
9. **for** (i=1; i<C, i++)
10.    **While** $R_i \in S$
11.    **for** $R_{[i]}$ to $R_{[i+1]}$
12.      Find total number of samples S
13.      Arr[$S_L$]= Left half of the total number of samples store in one array
14.      Arr[$S_R$]= Right half of the total number of samples store in another array
15.      S= Find minimum value of Arr[$S_L$]
16.      Q= Find minimum value of Arr[$S_R$]
17.      i++
18. **end**

Algorithm 2 is briefed as follows:

**Step 1:** Store R-peak value in an array.
**Step 2:** Find the minimum value between the first sample point to first R-peak value and mark it as $Q_{QRS}$ and store it in an array $Arr_{QRS}$[n]
**Step 3:** Now for the rest of each RR interval calculate the total number of samples and divide the total number of samples into two halves.
**Step 4**: Find the minimum value of each first half of the total number of samples and mark it S point of the QRS complex of ECG signal.
**Step 5**: Find the minimum value of each second half of the total number of samples and mark it S point of QRS complex of the ECG signal.
**Step 6**: Repeat step 4 and step 5 for rest of all RR intervals.

Once Q and S points are detected, the QRS value can be measured easily. Fig. 7 shows the QRS width of the ECG signal. Using the feature R-peak, we can easily calculate heart rate beat (HBR), shown in Fig. 8. Fig. 8 also shows heartbeat rate calculation and an effect of filtering of it. After QRS detection, the P wave is detected by the algorithm proposed by Hossain et. al.[30].

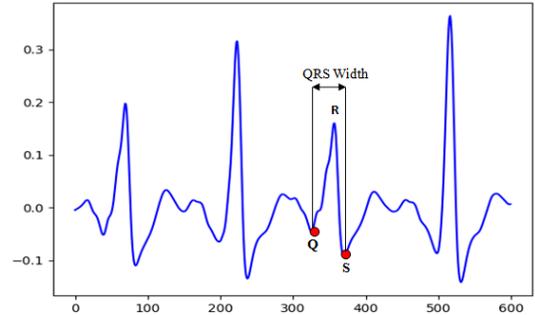

Fig 7: QRS width measurement of ECG signal

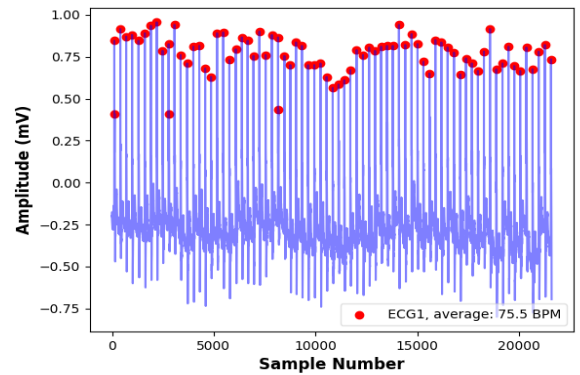

(a)

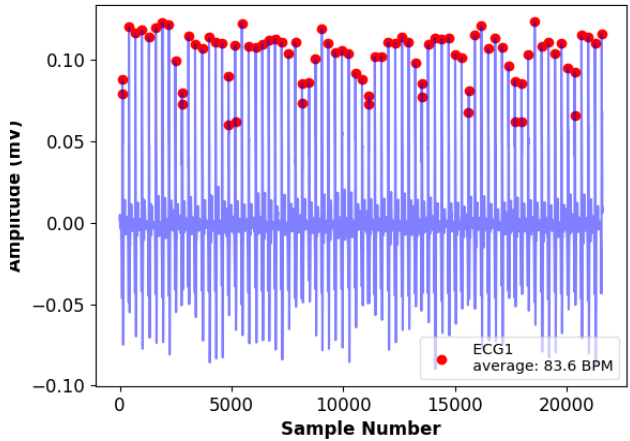

(b)

Fig 8: HBR calculation (a) before filtering (b) after filtering of record 101

*2.3 Classifier model:* We concentrate on correct ECG signal categorization in this section, which leads to the creation of a machine learning model based on supraventricular tachycardia arrhythmia diagnosis. Our main goal is to accurately identify the ECG sample beat and eliminate errors. As a result, it is crucial that critical signals are not misclassified as non-critical, which can lead to catastrophic device faults. Precision, F1 score as well as sensitivity, or recall are also used to estimate the performance of machine learning models such as decision tree, SVM, KNN, and logistic regression..

### III. SVT CLASSIFICATION BASED ON ECG FEATURES

There are four primary types of SVT: paroxysmal supraventricular tachycardia (PSVT), atrial fibrillation (AF), Wolff–Parkinson–White syndrome (WPW), and atrial flutter [14]. Among these four types, atrial fibrillation and Wolff-Parkinson-white syndrome are dangerous and are related to increased risk of cardiac arrest, heart failure, and stroke [15-16]. Based on the severity, we classify SVT that includes non-critical SVT, AF, and WPW.

*A. Atrial fibrillation with ECG findings*: Atrial fibrillation (AF) is an arrhythmia that is characterized by a fast, irregular heartbeat of the atrial chambers [32]. It generally starts with small episodes of aberrant beating, which over time becomes longer or constant [33]. It is the most frequent severe irregular rhythm of the heart and affects about 33 million individuals globally by 2020 [34]. Around 0.4% of females and 0.6% of men are affected in the developing world [35]. This disease has an increased risk of dementia, heart failure, and stroke. ECG features [36] of AF are shown in Table III.

*B. Wolff–Parkinson–White Syndrome with ECG findings:* Wolff–Parkinson–White syndrome is a disease caused by the symptoms of a specific type of cardiac electrical dysfunction [37]. Severe symptom includes cardiac arrest. Though it's rare but it may occur. It affects the population between 0.1 and 0.3%. The chance of mortality is around 0.5% annually in children and 0.1% annually in adults in individuals with no symptoms. ECG features [38] of WPW are shown in Table 3.

Table III. SVT Classification based on Features

| Features | Non-critical SVT | Atrial Fibrillation [39] | WPW [40] |
|---|---|---|---|
| **HBR range** | 100-250 BPM | 100-175 BPM | 160-300 BPM |
| **P wave** | Present | Not Present | Present |
| **PR interval** | 120 to 200 milliseconds | Not present | <120 ms |

### IV. RESULTS AND DISCUSSION

*A. Material*

ECG data from the PhysioNet database's MIT-BIH nsrdb, Cudb, and MIT-BIH svdb [40] are gathered and processed for this work in order to identify any differences in ECG signals. Each record of MIT-BIH svdb, MIT-BIH nsrdb, Cudb and MIT-BIH svdb, includes, 11730944 samples, 9205760 samples, 127232 and 230400 samples respectively. These sets reflect several subject categories as well as recording situations such as sampling speeds (128 and 250 Hz) and interferences. Every record's ECG1 data is utilized without exception.

*B. Simulation results:*

Python 3.7 is used in all simulations for electrocardiogram signal filtering as well as for training datasets and testing datasets. Tables IV and V demonstrate the results of several extracted characteristics with a small number of randomly picked signals before and after filtering. Table VI shows abnormalities in several signal parameters (IBI, SDNN, and RMSSD) identified in these experiments for certain chosen signals.

Using all ECG features described in section II, we prepared a machine learning-based SVT classification model. Using these features, we classify signals into three categories (atrial fibrillation, WPW, and non-critical SVT). ECG of record 04126 demonstrating atrial fibrillation is shown in Fig 9. ECG of record 203 demonstrating Wolff-Parkinson-white syndrome is shown in Fig 10. Fig 11 shows the non-critical SVT of ECG record 868.

Table IV. Results of various extracted features with a few number of random selected signals before filtering

| Features | Record 802 | Record 824 | Record 826 | Record 856 |
|---|---|---|---|---|
| RR interval (s) | 0.96 | 0.76 | 0.66 | 0.62 |
| QRS width (s) | Normal | Normal | Normal | Normal |
| HBR (BPM) | 63 | 79 | 92 | 95 |
| Supraventricular Tachycardia | no | no | no | no |

Table V. Results of various extracted features with a few numbers of randomly selected signals of Table I after filtering

| Features | Record 802 | Record 824 | Record 826 | Record 856 |
|---|---|---|---|---|
| RR interval (s) | 0.49 | 0.38 | 0.42 | 0.62 |
| QRS width (s) | Narrow | Narrow | Narrow | Normal |
| HBR (BPM) | 121 | 157 | 143 | 95 |
| Supraventricular Tachycardia | yes | yes | yes | no |

Table VI. Results of IBI, SDNN and RMSSD with a few number of records having SVT arrhythmia

| Record | IBI | SDNN | RMSSD |
|---|---|---|---|
| 16265 | 322.34 | 9.25 | 6.35 |
| 16272 | 269.86 | 131.80 | 217.88 |
| 16273 | 331.26 | 48.02 | 54.08 |
| 16420 | 165.29 | 60.13 | 109.92 |
| 16539 | 387.13 | 27.68 | 22.26 |

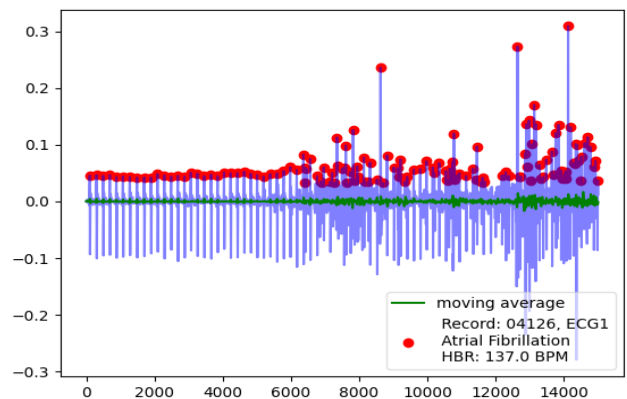

Fig 9: ECG of record 04126 demonstrating atrial fibrillation

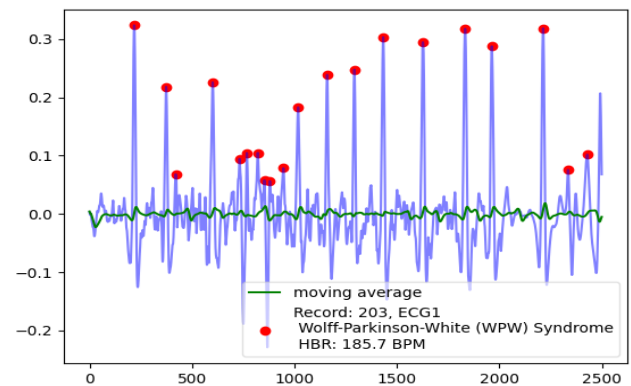

Fig 10: ECG of record 203 demonstrating wolff-parkinson-white syndrome

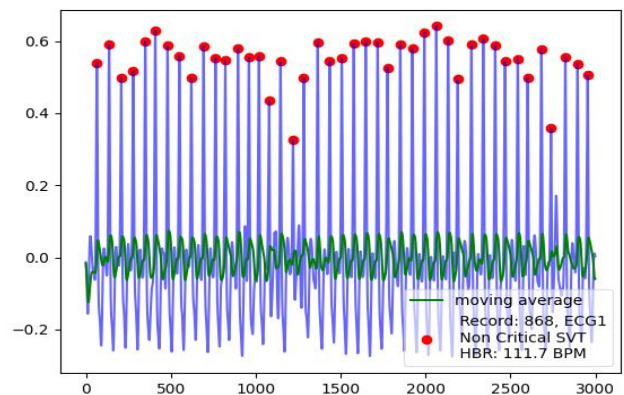

Fig 11: ECG of record 868 demonstrates non-critical SVT

Sensitivity, precision, F1 score and accuracy are measured as assessment of our work.
These definitions are given below.

1) The precision (PR): $PR = \dfrac{TP}{(TP + FP)}$

2) The sensitivity (SE): $SE = \dfrac{TP}{(TP + FN)}$

3) The specificity (SP) or F1 score: $SP = \dfrac{TN}{(TN + FN)}$

Where: FP = False Positives; TP = True Positives; FN = False Negatives; TN = True Negatives; and
N= FP + FN +TP +TN.

Table VII. Performance analysis

| Algorithm | Precision | Sensitivity | F1 Score | Accuracy (%) |
|---|---|---|---|---|
| KNN | 0.92 | 0.92 | 0.92 | 92 |
| SVM | 0.92 | 0.91 | 0.92 | 92 |
| Decision Tree | 0.95 | 0.95 | 0.95 | 95 |
| Logistic Regression | 0.94 | 0.94 | 0.94 | 94 |

Table VIII. Multiclass classification for decision tree based model

| Algorithm | Precision | Sensitivity | F1 Score |
|---|---|---|---|
| NCSVT | 0.89 | 0.94 | 0.91 |
| AF | 1.00 | 0.91 | 0.95 |
| WPW | 0.95 | 1.00 | 0.98 |
| Accuracy | | | 0.95 |
| macro avg | 0.95 | 0.95 | 0.95 |
| Weighted avg | 0.95 | 0.95 | 0.95 |

Precision is the percentage of an ECG signal that has previously been designated as critical. Sensitivity or recall refers to the fraction of the essential ECG signal designated as critical. In circumstances with unequal class distribution, the F1 score or specificity is the harmonic mean of precision and sensitivity, which is preferable to accuracy. In the instance of F1, Table VII shows that decision tree has the highest F1 score, followed by SVM, logistic regression, and KNN in decreasing order. Table VIII further shows that decision trees outperformed SVM, logistic regression, and KNN in terms of accuracy. As a result, a decision tree is superior for detecting crucial ECG signal bits. Multiclass classification among non-critical SVT (NCSVT), atrial fibrillation (AF) and Wolff–Parkinson–White syndrome (WPW) for decision tree based model is shown in table IX. In Table IX, our experimental results are compared with [26-28]. Experimental outcomes show satisfying improvements and great algorithm robustness that we have proposed.

Table IX. Comparative Study

| Author | Precision | Recall | F1-score | Accuracy |
|---|---|---|---|---|
| Zihlmann et al. [41] | - | - | 79% | 82% |
| Goodfellow et al. [42] | 84% | 85% | 85% | 88% |
| Jalali et al.[43] | 86% | 86% | 85% | 89% |
| Proposed method | 95% | 95% | 95% | 95% |

V. CONCLUSION

In this work, we suggested a machine learning approach to classify different types of supraventricular tachycardia arrhythmias of ECG signal. Our technique describes the pre-processing stage, feature extraction and classification model. In the pre-processing step, we de-trended and de-noised the signal to eliminate the noise for accurate function recognition. Afterwards, our proposed technique is used to detect R-peaks and QRS. The accompanying parameters, such as the PR interval, RR interval, QRS duration, HBR, RMSSD, SDSD, and IBI, are determined after this R-peak is noticed. We developed an effective machine-learning algorithm that can accurately classify various forms of supraventricular tachycardia based on these characteristics. This work effectively addresses the difficulty of reducing SVT misclassification. Experiment findings show that our proposed approach has evolved effectively and looks to be extremely stable, with 97 % accuracy.

To our knowledge, it is the first report that analyzes several models of machine learning and then selects a high-efficiency approach to classify different types of ECG supraventricular tachycardia arrhythmias. Our findings suggest that decision-tree-based models are the most effective model of machine learning for classification of suprarventricular tachycardia. With more advanced algorithms, we can learn to get a better outcome with even larger amount of data in the future as perfection has been highly anticipated in medical terms.

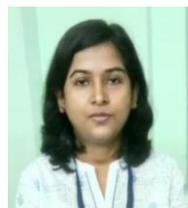

**Pampa Howladar** completed her B. Tech in Information technology from Govt. College of Engineering & Ceramic technology (GCECT) under MACAUT University. She received M.Tech degree in School of VLSI Technology from Indian Institute of Engineering Science and Technology, Shibpur (Formerly BESU). She received PhD degree in Information Technology in the area of Design and Optimization of Digital Microfluidic Biochips on Microelectrode-Dot-Array Architecture from

Indian Institute of Engineering Science and Technology, Shibpur, India in 2020. Her research interests are digital microfluidic biochips, Embedded Systems, Bioinformatics and Biomedical Signal Processing. Her research works appeared in reputed journals including IEEE transactions on Very Large Scale Integration (VLSI) Systems and IEEE/ACM Transactions on Computational Biology and Bioinformatics and also in referred international conference proceedings. She is now actively exploring machine learning and artificial intelligence techniques for biomedical applications. She has got best paper awards for her papers in Computing, Communication and Sensor Network Conference in 2018, International Conference on Microelectronics Circuit &Systems in 2021 and International conference on Data Analytics &Management in 2021. She also served as a Program Committee member of IEEE ISDCS, 2020.

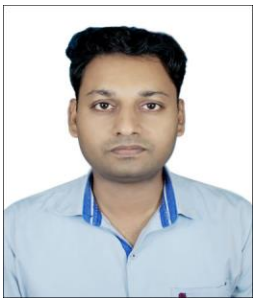

MANODIPAN SAHOO (M'14) was born in Haldia, West Bengal, India in 1983. He received M. Tech. in Instrument Technology from Indian Institute of Science, Bangalore in 2006. He received PhD degree from IIEST, Shibpur, India in 2016. His PhD thesis was on ``Modeling and Analysis of Carbon Nanotube and Graphene Nanoribbon based Interconnects". He is currently serving as an Assistant Professor in the Department of Electronics Engineering, Indian Institute of Technology (Indian School of Mines), Dhanbad, India. His research interests include Modeling and simulation of nano-interconnects and nano-devices, VLSI Circuits and Systems, Internet of Things, Biomedical Signal Processing . He has published more than 50 articles in archival journals and refereed conference proceedings. He has also published a Book entitled ``Modelling and Simulation of CNT and GNR Interconnects" with Lambert Academic Publishers in 2019. He published a book chapter entitled ``Modelling Interconnects for Future VLSI Circuit Applications" with IET Publishers in 2019. He is also associated as Member of IEEE, IETE, IEI and Life Member of Instrument Society of India.